# Unusual Pore Volume Dependence of Water Sorption in Monolithic Metal-Organic Framework

Jiawang Li[1#], Guang Wang[1#], Hongzhao Fan, Zhigang Li[1], Chi Yan Tso[2] and Yanguang Zhou[1*]

[1]*Department of Mechanical and Aerospace Engineering, The Hong Kong University of Science and Technology, Clear Water Bay, Kowloon, Hong Kong SAR*

[2]*School of Energy and Environment, City University of Hong Kong, Tat Chee Avenue, Kowloon Tong, Hong Kong SAR*

## Abstract

Monolithic metal-organic frameworks (MOFs), which have a continuous structure composed of small primary MOF particles and amorphous networks, are demonstrated to possess larger pore volume and thus better larger gas uptake capacity compared to their powder forms. Here, we systematically investigated the water vapor adsorption kinetics in a prototypical MOF, i.e., MOF-801. Our results show that the total pore volume (average pore diameter) of the monolithic MOF-801 is 0.831 $cm^3/g$ (5.20 nm) which is much larger than that of powder MOF-801, i.e., 0.488 $cm^3/g$ (1.95 nm). Unexpectedly, we find that the water uptake capacity of monolithic MOF-801 is much lower than that of powder MOF-801 when the RH ranges from 10% to 90%. Our molecular dynamics simulations further demonstrate that the unexpected water uptake capacity of monolithic MOF-801 at RH of 10%~90% is caused by the water film formed by the capillary condensation in these mesopores of monolithic MOF-801. The water molecules can overcome the capillary force when the RH is higher than 90%, and then leads to the increase of the corresponding water uptake capacity of monolithic MOF-801. Our findings reveal the underlying mechanisms for water adsorption kinetics in both powder and monolithic MOFs, which could motivate and benefit the new passive cooling or water harvesting system design based on MOFs.

---

[#] These authors contribute equally. [*]Author to whom all correspondence should be addressed. Email: maeygzhou@ust.hk

**INTRODUCTION**

Metal-organic frameworks (MOFs) are a class of porous materials consisting of metal clusters coordinated to organic ligands. They typically possess ultrahigh specific surface area[1,2], large porosity[3,4], and tunable pore size[5,6]. These outstanding properties make MOFs promising candidates for gas storage[7–9] and selective adsorption and separation[10–12]. MOFs are generally synthesized by the classical hydrothermal methods and in the form of powder particles. This loose state leads to poor mechanical properties[13,14] and durability[14,15], strongly affecting their applications. For example, nanoindentation testing using the atomic force microscope (AFM) tip revealed that Young's modulus of IRMOF-1 microcrystalline powder was only one-tenth of its theoretically predicted value[16]. Furthermore, in the application of flow reactors for atmospheric water capture, the reactor clogging or degradation of MOFs caused by the powder form will largely reduce the equipment's durability and service life[14]. While shaping MOFs through mechanical compression[17,18] and granulation[19,20] can improve their corresponding mechanical property and durability, the collapse of pores during mechanical compression[17] or the block of pores during granulation[20] led to the decreased porosity of MOFs.

Monolithic MOFs with hierarchical pores ranging from ~1.0 nm to ~20 $\mu$m have been demonstrated to possess excellent mechanical properties and durability[21,22] to tackle the above issues. For instance, the monolithic zeolitic imidazolate framework-8 (ZIF-8) showed much higher Young's modulus and hardness than its powder counterpart[16]. As another example, experiments show that the high bulk density of monolithic HKUST-1 largely improves its mechanical durability against permanent plastic deformation[7]. Meanwhile, the bulk density and pore volume of monolithic MOFs were generally much higher than those of their powder counterparts[7,8,16], resulting in a better capability for gas or water adsorption[7,21]. For instance, the methane uptake capacity of HKUST-1 at 65 bar and 298 K increases from 180 $cm^3/cm^3$ to a record-high value of 259 $cm^3/cm^3$ when its form changes from powder to monolith[7]. The monolithic HKUST-1 was the first qualified adsorbent for methane storage by the U.S. Department of Energy[23], and was further demonstrated to possess excellent hydrogen uptake capacity compared to that of powder HKUST-1[24]. Most recently, the MOF-801 has attracted much attention owing to its good water uptake capacity at ultralow relative humidity (RH, i.e., the water uptake is 0.25 g/g at an RH of 20% and 298 K) and high thermal stability[25,26], and therefore can be applied for water harvesting in arid area[26] and the development of passive cooling coatings[27,28]. The MOF-801 currently used in many applications is synthesized by the conventional hydrothermal method and therefore is in powder form. It is expected that its surface area and porosity can be further improved in its monolithic state, which can lead to higher water uptake capacity.

As previously discussed, despite the extensive literature on the synthesis and applications of powder MOF-801 and other monolithic MOFs, there is a noticeable scarcity of reports on their water adsorption mechanism. In this study, we systematically investigated the water adsorption kinetics in both powder and monolithic MOF-801. The as-prepared powder MOF-801 is formed by an aggregation of crystalline particles, whereas the monolithic sample has a continuous structure composed of small primary particles and amorphous networks. Our Brunauer-Emmett-Teller (BET) tests showed that the total pore volume (average pore diameter) of the powder and monolithic MOF-801 is 0.488 $cm^3/g$ (1.95 nm) and 0.831 $cm^3/g$ (5.20 nm), respectively. We further measured the corresponding water uptake capacities of powder and monolithic MOF-801 and found that the water uptake capacity of monolithic MOF-801 is unexpectedly much lower than that of powder MOF-801 in the RH ranging from 10% to 90%. When the RH is higher than 90%, the water uptake capacity of monolithic MOF-801 becomes larger than that of powder MOF-801 and increases dramatically from ~0.43 g/g at an RH of 90% to ~0.68 g/g at an RH of 97%. Our water adsorption isotherm tests and molecular dynamics (MD) simulations demonstrated that this unexpected lower water uptake of monolithic MOF-801 at RH of 10%~90% is caused by the water film formed by the capillary condensation in its mesopores. When the RH is higher than 90%, the higher water vapor partial pressure can then overcome the capillary force. Therefore, the corresponding water uptake capacity of monolithic MOF-801 increases dramatically.

**RESULTS AND DISCUSSION**

**Material synthesis and characterization**

The powder and monolithic MOF-801 are synthesized using the hydrothermal method[27] and sol-gel method[29], respectively. The schematic synthesis process is shown in **Figure 1a** (see **Methods** for details). Briefly, the fumaric acid, zirconium chloride oxide octahydrate ($ZrOCl_2 \cdot 8H_2O$), and the modulators were first dissolved in dimethylformamide (DMF) solvent to form a uniform solution by stirring, followed by the chemical reaction process. For synthesizing the MOF-801 powder, white precipitates are obtained after heating the stirred solution, and powder MOF-801 is then obtained after the vacuum filtration and activation of precipitates. For the synthesis of monolithic MOF-801, uniform and non-flowing amorphous gels are formed by heating the stirred solution, and centimeter-scale monolithic MOF-801 with bright and milky surfaces is then obtained through centrifugation, solvent exchange, and vacuum activation of the amorphous gels. The measured powder X-ray diffraction (PXRD) of the powder samples shows clear individual peaks and agrees well with the calculated PXRD[30] (**Figure 1e**) and previous results[27], which indicated the high crystallinity of our synthesized samples (**Figure S7**). Whereas, even though the peak position of monolithic MOF-801 is the same as powder MOF-801, the peaks of monolithic MOF-801 are much broader which indicates a lower crystallinity than the powder MOF-801, which is confirmed by the following micromorphology analysis.

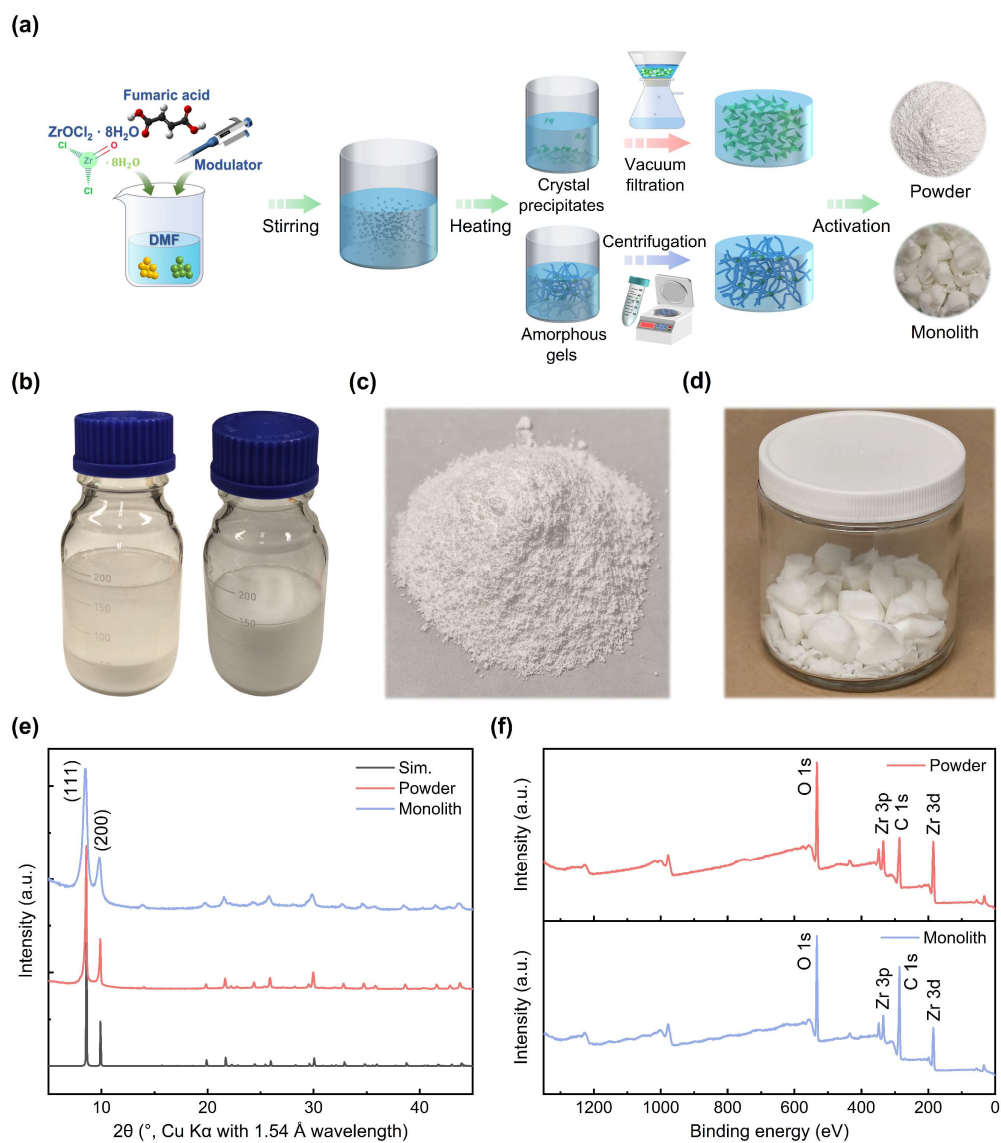

**Figure 1 The synthesis and characterizations of powder and monolithic MOF-801.** (a) The schematic of the synthesis process of powder and monolithic MOF-801. (b) The crystal precipitates during the synthesis of the powder MOF-801 after heating (left), and the amorphous gel during the synthesis of the monolith MOF-801 after heating (right). (c) The powder MOF-801 after activation and grinding. (d) The monolithic MOF-801 after activation. (e) The PXRD patterns of powder and monolithic MOF-801. The XPS for (f) powder MOF-801 and monolithic MOF-801.

We further characterized the chemical structure and functional group information of powder and monolithic MOF-801 through Raman spectrum analysis and Fourier transform infrared (FTIR) measurements. Our Raman (**Figure S2**) and FTIR (**Figure S3**) results are consistent with previously measured results[27,31,32] (see **Supplementary Note 1** for details), which indicates the pure phase of MOF-801. The X-ray photoelectron spectroscopy (XPS) (**Figures 1f**) is used to verify the element valence state and chemical composition of powder and monolithic MOF-801 (see **Supplementary Note 1 and Figure S4** for details), and confirmed the presence of C, O, and Zr elements in our synthesized powder and monolithic MOF-801. We also analyzed the thermal properties of MOF-801 by thermogravimetric analysis (TGA), and both the powder and monolithic MOF-801 showed similar weight loss curves and experienced four stages until the final solid product formed as $ZrO_2$ (see **Supplementary Note 1 and Figure S5** for details), which agreed with previous results[27,33,34].

**Micromorphology and pores in MOF-801**

The micromorphology of powder and monolithic MOF-801 was further investigated by the scanning electron microscope (SEM) and transmission electron microscope (TEM). Our results show that the powder MOF-801 exhibits crystalline particles which are distributed uniformly and have an average grain size of ~320 nm (**Figure 2a-c**), while monolithic MOF-801 shows a continuum form of gels with polycrystalline and amorphous phases (**Figure 2d-f** and **Figure S6**). The energy dispersive spectrometer (EDS) (**Figures 2g and h**) analysis is further used to verify the chemical composition of powder and monolithic MOF-801 (see **Supplementary Note 1 and Figure S4** for details) and confirmed the uniform distribution of C, O, and Zr elements in our synthesized MOF-801.

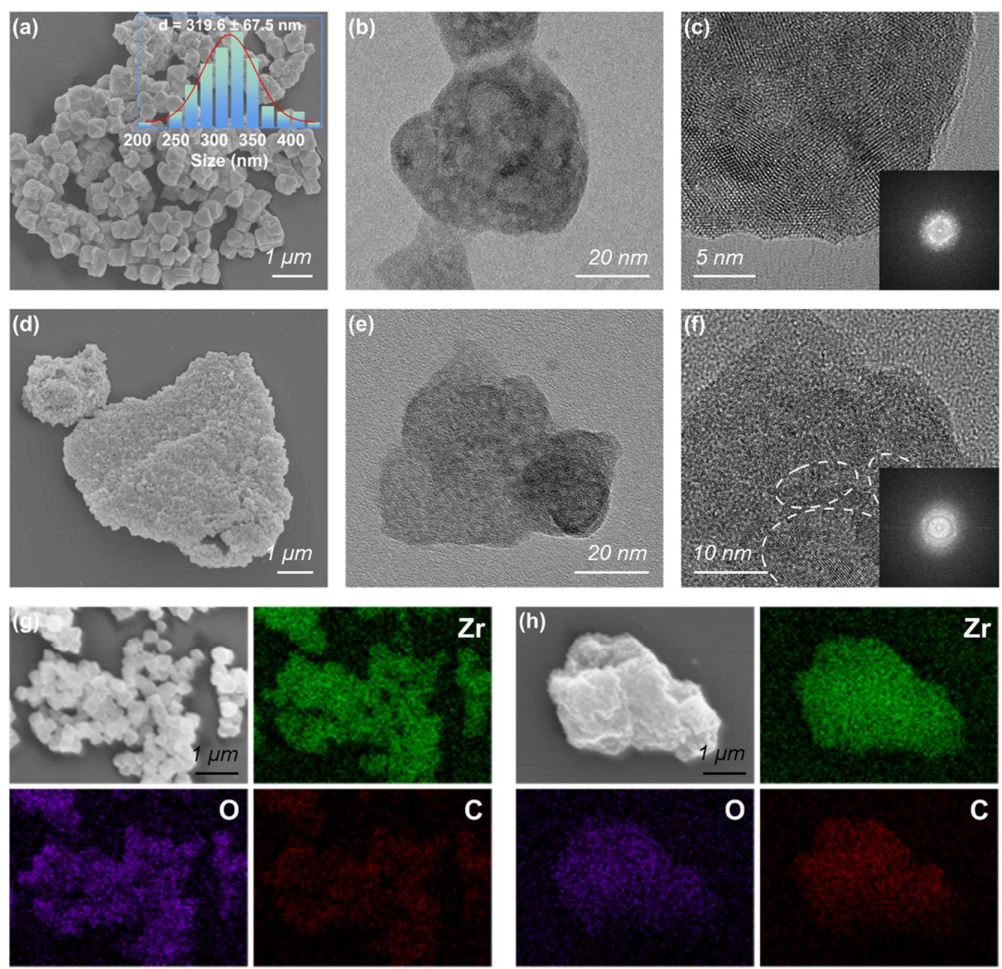

**Figure 2 The SEM, TEM, and EDS images of powder and monolithic MOF-801.** The morphology and the Fast Fourier transformation (FFT) spots for (a, b, c) the powder MOF-801 and (d, e, f) the monolithic MOF-801. The elemental mapping of Zr, O, and C of the regions for (g) the powder MOF-801 and (h) the monolithic MOF-801.

We calculated the crystallinity (**Figure S7**) of samples from the XRD patterns using the XRD deconvolution method in MDI Jade software[35]. The crystallinity of the powder and monolithic MOF-801 obtained are ~91.2% and ~71.4%, respectively, which confirmed the different phase structures observed in TEM images. This difference can result in a large discrepancy in the physical properties we focused on for water adsorption applications, such as specific surface area, pore size distribution, and pore volume (**Figure 3a, Figure S8, and Table S1**).

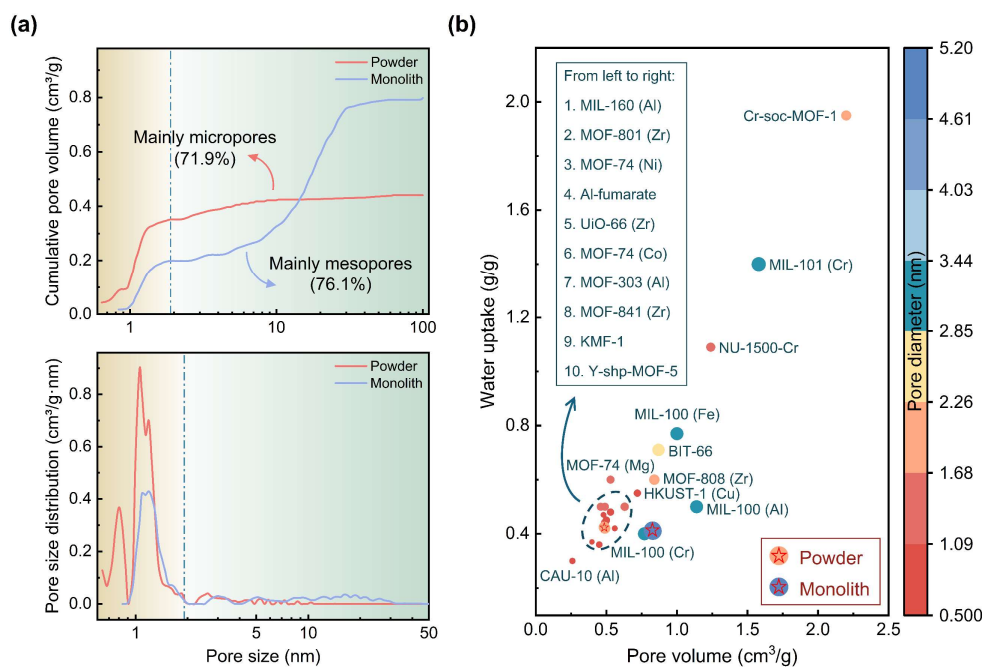

**Figure 3 The water uptake capacity and pore size distribution of powder and monolithic MOF-801.** (a) The Non-Local Density Functional Theory (NLDFT) cumulative pore volume and pore size distribution of powder and monolithic MOF-801 as a function of pore size. (b) The water uptake of 21 types of classical MOFs at 298K and an RH of 90%~95% as a function of the pore volume.

The $N_2$ adsorption-desorption isotherm relation obtained by Brunauer-Emmett-Teller (BET) tests indicates that the pore size of powder MOF-801 ranges from ~0.5 nm to ~10 nm, and the pore size of monolithic MOF-801 is in the range from ~1 nm to ~30 nm (**Figure 3a**). For the powder MOF-801, these intrinsic micropores[36] with diameter smaller than 2 nm contribute to ~71.9% of the total pore volume with a value of 0.488 cm$^3$/g (**Figure 3a**). The powder MOF-801 possesses a high specific surface area of 1002.3 m$^2$/g, and an average pore diameter of ~1.95 nm. For monolithic MOF-801, the intrinsic micropores contribute ~23.9% to the total pore volume with a value of 0.831 cm$^3$/g, and the rest are mesopores[36] with diameters between ~2 nm to ~50 nm (**Figure 3a**). The specific surface area of monolithic MOF-801 is 639.8 m$^2$/g which is lower than that of powder MOF-801, and an average diameter of ~5.20 nm which is much larger than that of powder MOF-801. The density and porosity of powder and monolithic MOF-801 are further measured by mercury intrusion porosimeter (MIP) tests (**Table S1**). Our MIP results showed that the bulk densities of powder and monolithic MOF-801 are 0.478 cm$^3$/g and 0.815 cm$^3$/g, and the porosity of powder and monolithic MOF-801 is 64.2% and 58.0%, respectively.

**Water adsorption capacity**

We next systematically investigated the water uptake capacity of both powder and monolithic MOF-801, and all the water uptake experiments were carried out at room temperature. For comparison, we summarized the physical properties and corresponding water uptake performance of 21 types of classical MOFs at RH of 90% or 95% in **Figure 3b** (see **Table S2** for details), including their pore diameter, pore volume, and easily found the obvious linear relationship between the pore volume and water uptake capacity. As discussed above, the total pore volume of monolithic MOF-801 is almost double that of powder MOF-801. Therefore, it is expected that the monolithic MOF-801 will have a much higher water uptake capacity per unit mass than the powder MOF-801. We then studied water adsorption isotherm curves using BET to measure the water uptake (**Figure 4a**). For the powder MOF-801, it is found that the water vapor begins to be absorbed dramatically to a plateau value of ~0.25 g/g when the water vapor pressure reaches the critical pressure (i.e., the RH is ~10% when the water vapor pressure is ~317 Pa)[25,30]. When the vapor pressure is larger than the critical pressure, the water vapor uptake capacity increases gradually with the increase of the vapor pressure (**Figure 4a**), and the water uptake capacity of powder MOF-801 can be as high as ~ 0.42 g/g at an RH of 90%. For the monolithic MOF-801, both micropores with diameters smaller than 2 nm and mesopores with diameters ranging from 2 nm to 30 nm coexist as discussed above. The water vapor is gradually adsorbed by the monolithic MOF-801 and reaches a value of ~0.34 g/g at the RH of 80%. When the RH is higher than 80%, the water uptake capacity increases dramatically to a high value of ~0.68 g/g (i.e., at which the RH is 97%). Meanwhile, it is worth noting that the water uptake capacity of monolithic MOF-801 is lower than that of powder MOF-801 when the RH is in the range of 10%−90%. This is caused by the capillary condensation of water vapor on the surfaces of mesopores (detailed discussion can be found below).

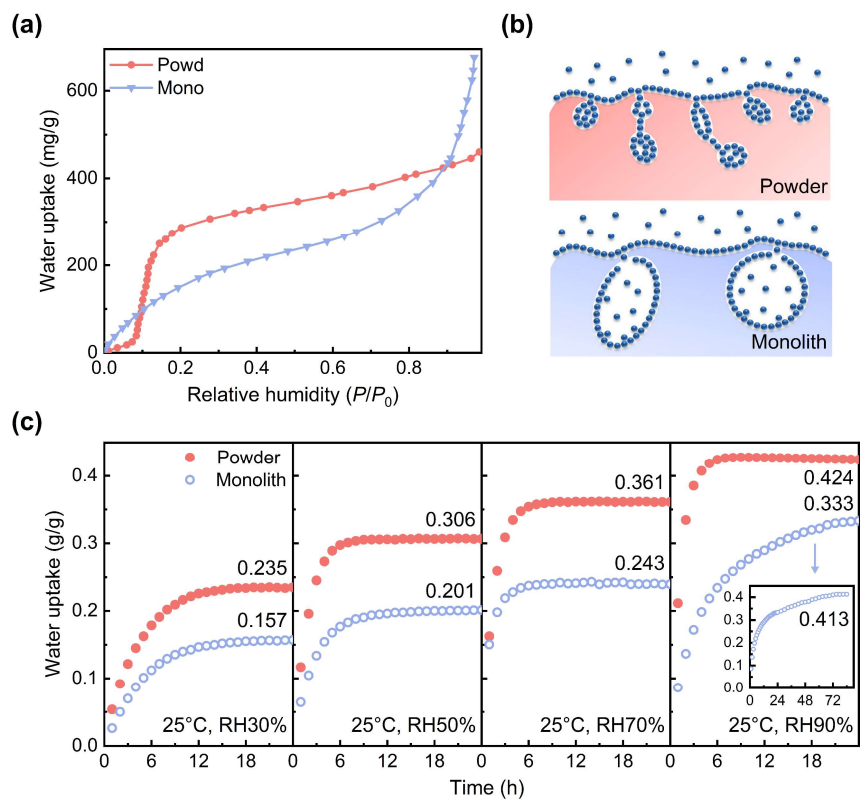

**Figure 4 The water uptake of powder and monolithic MOF-801.** (a) The water vapor adsorption isotherms of powder and monolithic MOF-801. (b) The schematic of water adsorption mechanisms in powder and monolithic MOF-801. (c) The on-site water uptake tests of powder and monolithic MOF-801 at 25 °C and RHs of 30%, 50%, 70%, and 90%.

We also conducted on-site water adsorption tests to measure the water uptake capacity in real application scenarios (see **Methods** for details). For the RH of 30%, 50%, and 70%, all the on-site water adsorption tests are implemented in the control room (**Figure S9a**). For the RH of 90%, on-site water adsorption tests are performed using our home-made chamber with constant temperature and humidity (**Figure S9b**). To be more consistent with the practical applications and reduce measurement errors, up to 50 g powder or monolithic MOF-801 was used in our on-site tests. Our results showed that the saturation time of the adsorbed water in both powder and monolithic MOF-801 decreased with the increased RH. For instance, the saturation time for powder MOF-801 is ~12 and ~6 hours at the RH of 30% and 70%, respectively. This is because when the vapor pressure increases with the RH, more vapors can be easily adsorbed in these pores. At RHs of 30%, 50%, and 70% (**Figure 4c**), the water uptake is 0.235 g/g, 0.306 g/g, and 0.361g/g, respectively, which are higher than those of monolithic MOF-801 (i.e., 0.157 g/g, 0.201 g/g and 0.243 g/g, respectively). When the RH is 90%, the water uptake of monolithic MOF-801 is 0.413 g/g which is almost equal to that of powder MOF-801 (i.e., 0.424 g/g). It is noted that all the water uptake in the on-site water adsorption tests is lower than the corresponding value in the water adsorption isotherm tests. For example, the water uptake of powder and monolithic MOF-801 is ~0.31 g/g and ~0.19 g/g by the BET tests at the RH of 30%, but the counterpart is 0.235 g/g and 0.157 g/g in the on-site water adsorption tests. This may be because the sample in the BET test is only about 0.1 grams, which occupies a small space in the test chamber, allowing the sample to have more surfaces exposed to water vapor, and this is consistent with previous reports[37].

**Underlying mechanisms for the water adsorption process**

We then investigated the underlying mechanism that results in the difference of water adsorption capacity between powder and monolithic MOF-801. The wettability measurement showed that the contact angles for powder and monolithic MOF-801 are ~47° and ~37° (see **Supplementary Note 2** and **Figure S11** for details), respectively. Therefore, both powder and monolithic MOF-801 are hydrophilic, and the water drops can easily spread on their surfaces. This hydrophilicity can facilitate water adsorption in MOFs.

It has been suggested that the water vapor or some organic solvent vapor adsorption capacity of porous crystals is mainly determined by the specific surface area at low RHs and by the pore volume at high RHs[38,39]. The reason is that[38–40], at low RH, the molecules are primarily adsorbed to the adsorption sites on the inner surface. Thus, the adsorption capacity is related to the number of adsorption sites which is related to the specific surface area. At high RH, pore filling occurs and the larger pore volume provides more space for molecules. Our BET tests showed that the pores in powder MOF-801 are mainly micropores with pore sizes of ~0.7 nm and ~1.2 nm (**Figure 3b**). For monolithic MOF-801, ~23.9% of the pores are micropores and the rest are mesopores with sizes of ~2 – ~50 nm (**Figure 3b**). The large number of mesopores in monolithic MOF-801 will lead to a higher pore volume but a lower specific surface area compared to that of powder MOF-801. Our results showed that the specific surface area for powder MOF-801 is 1002.3 m$^2$/g and much larger than that of monolithic MOF-801 (i.e., 639.8 m$^2$/g). The pore volume of powder and monolithic MOF-801 are 0.488 cm$^3$/g and 0.831 cm$^3$/g, respectively. As a result, the powder MOF-801 possesses a much higher water adsorption capacity compared to that of monolithic MOF-801 when the RH is smaller than 90%. When the RH is higher than 90%, the water adsorption capacity of monolithic MOF-801 becomes larger than that of powder MOF-801.

We also studied the water adsorption process in powder and monolithic MOF-801 using molecular dynamics (MD) simulations. Our MD simulations show that water molecules are uniformly adsorbed in the intrinsic micropores (**Figure 5a**). For these mesoscale pores, water molecules will form a water film on the surface of mesoscale pores due to the surface capillary effect (**Figures 5b and 5c**), which is so-called capillary condensation[41,42]. When the water vapor pressure is high enough to break the formed water film (i.e., the corresponding RH is higher than 90% at 25 °C for monolithic MOF-801), water molecules can then be further adsorbed by these mesoscale pores. Therefore, the water uptake capacity of monolithic MOF-801 becomes higher than that of powder MOF-801 when the RH is higher than 90% at 25 °C.

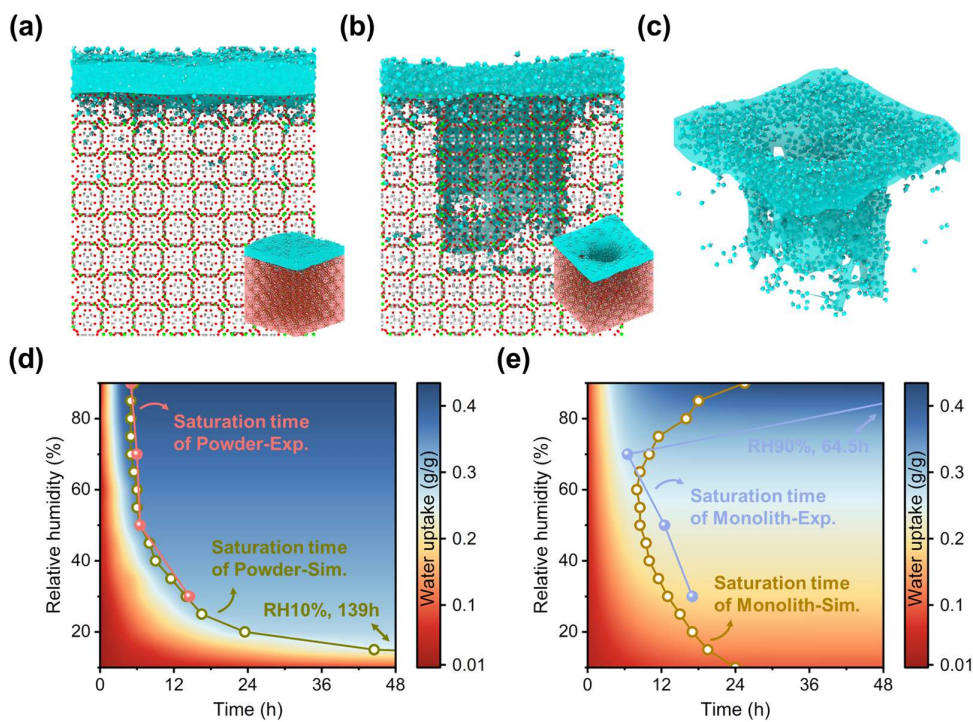

**Figure 5 The simulations of water adsorption of powder and monolithic MOF-801.** The MD simulations of (a) atomistic micropores and (b) atomistic mesopores on water adsorption of MOF-801. (c) The water film in the mesopores of monolithic MOF-801. The FEM simulations of water adsorption isotherm spectrum and saturation time of (d) powder MOF-801 and (e) monolithic MOF-801, the solid circles and open circles are the saturation times of MOF-801 obtained in experiments and simulations, respectively.

**Water adsorption isotherm spectrum**

Recently, many applications based on the water adsorption-desorption of MOFs have been proposed, such as cooling[27,43], water harvesting[25,26,44], and humidity control[45,46]. Their corresponding performance is determined by RH, water saturation time, and water uptake. To guide the design for real applications, we systematically mapped the water adsorption isotherm spectrum considering RH and water saturation time of both powder and monolithic MOF-801 using finite element method (FEM) simulations. The FEM model is constructed in the COMSOL Multiphysics software, in which mass diffusion and moisture adsorption equations are coupled (details can be found in the **Methods and Supplementary Note 3**). The saturation time of the water adsorption is determined as the time used to achieve 98% of the total water uptake. It is found that the saturation time calculated in our FEM simulations agree well with our experimental measurements (**Figure 5d and e, and Figure S13**), which indicates our FEM simulations capture the real water adsorption process in MOF-801 accurately.

Our FEM simulations further showed that the water uptake of powder MOF-801 converged to the saturated values (i.e., from 0.315 to 0.419 g/g at RH of 35% ~ 90%) in a short time of less than 12 hours when the RH is higher than 35% (**Figure 5d**). At a high RH larger than 55%, the saturation time is only 5 ~ 6 hours (**Figure 5d**). The fast water adsorption process indicates the potential applications of powder MOF-801 in passive evaporative cooling technology, which utilizes the day-and-night water uptake difference of MOF-801[47,48]. The theoretical average cooling power of the powder MOF-801 at 25 °C ranges from 18.4 to 125.8 W/m$^2$ when the RH changes from 20% to 90%, in which we assume all the adsorbed water evaporates in the corresponding saturated time. For the monolithic MOF-801, the corresponding water uptake converges to the saturated values (i.e., from 0.198 to 0.308 g/g) in 12 hours when the RH is at 35 ~ 75% (**Figure 5e**). The theoretical average cooling power of the monolithic MOF-801 at 25 °C ranges from 13.8 to 48.9 W/m$^2$ when the RH changes from 20% to 80%. It is noted that during the water desorption process, the desorption speed depends on the temperature of the substrate; thus, the maximum cooling power is much larger than the average cooling power. For instance, the maximum cooling power is 97.1 W/m$^2$ for monolithic and 166.2 W/m$^2$ for powder based on the adsorption curves. However, it takes more than 10 hours for monolithic MOF-801 to adsorb water to saturation at RHs larger than 80%. Therefore, while monolithic MOF-801 has almost equal water uptake (i.e., 0.413 g/g) at the RH of 90% compared to that of powder MOF-801 (i.e., 0.424g/g), it is not recommended to use monolithic MOF-801 for continuous passive cooling or water harvesting applications due to the longer saturation time. However, the monolithic MOF-801 can be used in moisture control, especially in the high moisture space. Moreover, the centimeter size of the monolith tackled the issue of powder form that is difficult to collect or easily blocks adsorption beds. These merits can facilitate MOF-801's industrial production, increase the lifetime of adsorption beds, and also reduce maintenance costs. This research can guide the MOF's synthesis and design for their real applications based on water adsorption.

**CONCLUSIONS**

In conclusion, we have systematically investigated the water adsorption kinetics of both powder and monolithic MOF-801. The as-prepared powder MOF-801 is composed of aggregated crystal particles, while the monolithic sample consists of a continuous structure made up of small primary particles and amorphous networks. Our BET tests showed that the pore volume (average pore diameter) of the powder and monolithic MOF-801 is 0.488 cm$^3$/g (1.95 nm) and 0.831 cm$^3$/g (5.20 nm), respectively. We further measured the corresponding water uptake capacities of powder and monolithic MOF-801, and we found that the water uptake capacity of monolithic MOF-801 is unexpectedly much lower than that of powder MOF-801 when the relative humidity ranges from 10% to 90%. When the RH is higher than 90%, the water uptake capacity of monolithic MOF-801 becomes larger than that of powder

MOF-801 and increases dramatically from ~0.43 g/g at an RH of 90% to ~0.68 g/g at an RH of 97%. Our water adsorption isotherm tests and molecular dynamics simulations demonstrated that the unexpected water uptake capacity of monolithic MOF-801 at RH of 10%~90% is caused by the capillary condensation in these mesopores of monolithic MOF-801. When the RH is higher than 90%, the water molecules can then overcome the hindrance from the capillary condensation and be adsorbed. The corresponding water uptake capacity of monolithic MOF-801 therefore increases dramatically. This provides invaluable guidance for the large-scale synthesis of MOFs and their utilization in water adsorption. Furthermore, these insights could facilitate the exploration of innovative applications such as passive cooling or water harvesting systems based on MOFs.

**Data and code availability**

The data and code that support the findings of this study are available from the corresponding author upon request.

**METHODS**

**Materials synthesis of MOF-801**

All the chemicals in our experiments, including fumaric acid ($C_4H_4O_4$, Macklin, 99.0%), zirconium chloride oxide octahydrate ($ZrOCl_2·8H_2O$, Aladdin, 98%), dimethylformamide (DMF) ($HCON(CH_3)_2$, RCI Labscan, 99.9%), formic acid (HCOOH, Scharlab, 98-100%), glacial acetic acid ($CH_3COOH$, Aladdin, ≥ 99.8%), concentrated hydrochloric acid (HCl, Scharlab, 37%), were used as received without any additional treatment.

The powder MOF-801 was synthesized referring to a conventional hydrothermal method[27]. Fumaric acid (1.74 g, 15 mmol) and $ZrOCl_2·8H_2O$ (4.83 g, 15 mmol) were accurately weighed and dissolved in 60 ml of DMF. Formic acid (20 ml) was added as the modulator. The resulting mixture was transferred to a 250 ml beaker and magnetically stirred for an hour at room temperature until a uniform solution was formed. The solution was then sealed in a 250 ml Schott bottle and heated at 130 °C for 6 hours in an oven. After cooling down to room temperature, white precipitates were clearly shown in the supernatant. The supernatant was poured out, and the white precipitate was collected using vacuum filtration with a 0.45 μm Nylon membrane filter (Tianjin Jinteng Experiment Equipment Co., ltd.). The collected solid was washed three times with deionized (DI) water and then dried in a vacuum oven at 150 °C for 24 hours to activate the synthesized MOF-801 powder sample.

The monolithic MOF-801 was synthesized based on a method for synthesizing hierarchical-porous monolithic material reported in the literature[29] with some modifications. The first step is the formation of the amorphous gel. In detail, fumaric acid (1.16 g, 10 mmol) and $ZrOCl_2·8H_2O$ (3.22 g, 10 mmol) were accurately weighed and dissolved in 60 ml of DMF.

Glacial acetic acid (2.0 ml) and concentrated hydrochloric acid (1.5 ml) were added as modulators to adjust the chemical reaction conditions. The resulting mixture was transferred to a 250 ml beaker and magnetically stirred for an hour at room temperature until a uniform solution was formed. The solution was then sealed in a 250 ml Schott bottle and heated at 100 °C for 6 hours in an oven. After cooling down to room temperature, a milky gel state was observed, indicating the formation of the gel. If the Schott bottle was inverted, and the amorphous gel was found to adhere well to the bottom of the bottle, rather than flowing down. The following step is to wash the gel through solvent exchange, the resulting gel and 50 ml DMF were mixed in a beaker and magnetically stirred for 30 minutes until the gel was evenly dispersed. The mixture was then transferred to a centrifuge tube and centrifuged at 5500 rpm for 3 minutes. The supernatant was poured out, followed by a similar gel washing and centrifugation (5500 rpm, 180 min). The centrifuged gel was then placed in a chamber with constant temperature and humidity (25 °C, 30% RH) overnight, and finally activated by heating at 100 °C in a vacuum oven for 12 hours.

The study also explored the scalable production of monolithic MOF-801. Through some modifications, including the proportional amplification of chemicals' quantity, the improvement of the experimental environment (i.e., extended time of heating, drying, and centrifugation), and adjustment of experimental parameters (i.e., reduced solvent usage), we can improve the mass scale from 2.4 g per batch to 50 g per batch (**Figure S3c**). This output not only met the requirements for more extensive and accurate characterizations of the materials but also paved the way for the potential industrialization of the monolithic MOF-801.

**Characterizations of MOF-801**

Samples for PXRD analysis were prepared by thoroughly grinding the sample with an agate mortar. The ground sample was then placed on a silica wafer in a shallow groove. The PXRD data of MOF-801 was characterized using a high-resolution powder X-ray diffractometer (Panalytical, X'pert Pro, Netherlands) with Cu Kα ($\lambda$ = 1.5406 Å, 30 kV) radiation. The measurements were carried out in the 2θ range from 5° to 45°, with a 0.033° increment and an accumulation time of 3 seconds at each point. The XRD data was analyzed using MDI Jade software version 6 to calculate the degree of crystallinity of samples. The XRD profiles were also fitted with pseudo-Voigt functions. The surface micromorphology and lattice pattern of the samples were characterized using the scanning electron microscope (SEM, JEOL JSM-6700F, Japan) and transmission electron microscope (TEM, JEOL JEM-2010, Japan). The microscopy images were processed using Image J software for analysis. The energy Dispersive Spectrometer (EDS) coupled with the SEM was used to perform quantitative and qualitative analysis of the elemental composition of the samples.

Meanwhile, the samples were also finely ground in an agate mortar and then dispersed in DI water using ultrasound. The resulting suspension was then dropped on the SEM sample stage or TEM grid and dried in air for further measurements. Fourier transform infrared spectroscopy (FTIR, Bruker, Vertex 70 Hyperion 1000, Germany) was used to obtain functional group information about the samples. The Raman spectrum was measured using a Raman spectrometer (Raman, RAMANMICRO 300, USA) with a laser wavelength of 785 nm, which was used to determine the vibrational and rotational energy levels of the molecules. All Raman spectra were measured directly on the solid sample with laser powers of 20 mW. The chemical composition and valence state of the samples were analyzed using X-ray photoelectron spectroscopy (XPS, Thermo Scientific K-Alpha, USA). The thermal properties of the samples were analyzed using TGA measurement (TA, Q5000, USA) in an $N_2$ atmosphere. The measurements were conducted in a temperature range from 20 °C to 800 °C at a heating rate of 5 °C/min. The contact angle meter (Biolin, Theta Flex, Finland) was used to measure the contact angle of the MOF tablets which could tell us the hydrophilicity of the samples. To prepare the tablets, we first placed an equal mass of powder or monolith into a mortar and ground it thoroughly. Then the hydraulic press (320 MPa for 5 mins) was applied to create a round tablet with a 12 mm diameter.

Furthermore, $N_2$ isotherms (adsorption and desorption) were collected using a surface area and pore size distribution analyzer (Micromeritics 3Flex, USA) at 77 K. The specific surface area is obtained through the BET model analysis, and the pore size distribution was obtained from adsorption isotherms using the Non-Local Density Functional Theory (NLDFT) method with the slit pore model. Before conducting the tests, in situ degassing (130 °C, 12 h) was performed under vacuum to ensure that all the water molecules inside the pores of MOF were completely evacuated. The density of samples (including bulk density and apparent density) at atmospheric pressure was determined using a Mercury intrusion porosimeter (MIP, MicroActive AutoPore V 9600, Micromeritics, USA), with the final pressure up to 60000 psi, and the analyzable pore sizes ranging from 3 nm to 800 μm. The samples were degassed overnight under vacuum at 100 °C before the tests. The water adsorption isotherms characterize the water adsorption performance of the samples. The water adsorption isotherms were collected using the same measurement system as the $N_2$ isotherms, and the only difference is the adsorbed gas was replaced by water vapor.

**Water adsorption performance in field tests**

The water absorption performance of MOF-801 was measured at various relative humidities in a control room (Jockey Club Controlled-Environmental Test Facility, HKUST) and a home-made chamber with constant temperature and humidity. Before the measurements, samples were dried in a vacuum oven at 100 °C for 12 hours to remove the adsorbed water. The mass change was recorded using an electronic balance (OHAUS, PR223ZH/E, USA) with an accuracy of 0.001 g, and temperature and humidity changes in the sealed container were recorded using a thermometer. The experiments at RHs of 30%, 50%, and 70% were carried out in the control room at 25 °C, the measurement error for relative humidity and temperature were ± 2% and ± 0.5 °C, respectively. The experiment at RH of 90% was conducted in a home-made chamber where the relative humidity was adjusted by a humidifier and controlled by a humidity sensor. The measurement error for relative humidity and temperature were ± 5% and ± 2.0 °C.

**Finite element method (FEM) simulations**

COMSOL Multiphysics software platform was utilized to establish a numerical model using FEM to investigate the water vapor adsorption behavior of porous MOF-801. Modules of laminar flow and moisture transport in porous media were chosen. Based on the experimental settings of on-site tests, a two-dimensional numerical model of the water vapor adsorption in the porous medium (i.e., MOF-801 layer) was established, as illustrated in **Figure S12**. The closed space and the porous medium in the space were selected as the research objects, the cross-sectional dimensions of the closed space are 500×20 mm$^2$ and the porous medium is 200×1.0 mm$^2$, and they are divided into 33581 meshes. The model simulates the process of laminar moist air flowing from left to right through the closed space and the porous medium. The porous medium can adsorb moisture from the atmosphere until it reaches saturation. The material properties and environmental condition parameters are summarized in **Table S3**. The transport and adsorption mechanism of water vapor in porous media is solved by relevant partial differential equations (PDEs) based on the principle of mass conservation.

In the simulation, we assumed that the liquid phase and the gas phase were in equilibrium inside the porous media, and neglected the influence of gravity on transport. The schematic diagram of the model is shown in **Figure S12**. In the domain of humid air, it is assumed that there is no liquid water present, meaning that all water enters the porous medium in the vapor state. The flow of moist air is driven by both pressure gradient $p_g$ and capillary pressure $p_c$, The velocity field $\boldsymbol{u}_g$ and pressure gradient $p_g$ of moist air can be calculated using the Brinkman equation. The water velocity $\boldsymbol{u}_l$ is calculated based on Darcy's law (1), which is defined by the pressure gradient in the gas phase. Referring to previous research reports[49], based on the summation of the mass conservation equations of water vapor and liquid water, the following equation (2) is derived to calculate the change of humidity $w(\varphi_w)$ in porous media with time

$$\boldsymbol{u}_l = -\frac{\kappa_l}{\mu_l}\nabla p_g \tag{1}$$

$$\frac{\partial w(\varphi_w)}{\partial t} + \rho_g \boldsymbol{u}_g \cdot \nabla \omega_v + \boldsymbol{u}_l \cdot \nabla \rho_l + \nabla \cdot (-\rho_g D_{eff} \nabla \omega_v) + \nabla \cdot (-D \frac{\partial w(\varphi_w)}{\partial \varphi_w} \nabla \varphi_w) = 0 \tag{2}$$

$$D_{eff} = D_{va} \varepsilon_p^{4/3} s_g^{10/3} \tag{3}$$

where, $\kappa_l$ and $\mu_l$ are the permeability and viscosity of the liquid phase, respectively. The total humidity comprises water vapor and liquid water in porous media, defining the saturation variables Sg and Sl of the gas phase (moist air) and liquid phase (water), which satisfy the constraint that their sum is 1. During the calculation process, the effective diffusivity $D_{eff}$ is obtained by relating the diffusivity in porous media to the Millington & Quirk equation (3), considering the porosity $\varepsilon_p$ of the porous media and the mass fraction $\omega_v$ of water vapor. The values of water vapor diffusivity $D_{va}$ and permeability $\kappa$ refer to previous reports[49,50]. For qualitative analysis, the environment and water temperature are assumed to be set at 25°C, and the parameters related to water are directly from the material library of COMSOL Multiphysics.

**Molecular dynamics simulations**

All the MD simulations are implemented by the Large-scale Atomic/Molecular Massively Parallel Simulator (LAMMPS)[51]. In MD simulations, the size of the MOF-801 model is 71.3 × 71.3 × 71.3 Å$^3$. To mimic the mesoporous in monolithic MOF-801, a pore with a volume of 35.6 × 35.6 × 53.5 Å$^3$ is constructed in the pristine MOF-801. The mesoscale pore starts from the top surface of the MOF-801 model. The length of the mesoscale pore along three directions is several times the lattice constant of MOF-801 to ensure the model charge is neutral. The atomic charge of Zr, O (connect with Zr and C), C (connect with O and C), C (connect with C and H), H, and O (only connect with Zr) in MOF-801 are 1.89, -0.537, -0.53, 0.59, -0.10, 0.16, and -0.792, respectively. The water model of the flexible variant of the simple point charge model (SPC/Fw)[52] was selected to describe the water molecules. The interactions between atoms in MOF-801 and water molecules were extracted from the universal force field (UFF)[53]. The cut-off distance of the van der Waals interactions is 12 Å. The long-range electrostatic interactions were considered by the particle-particle particle-mesh (PPPM) solver with a relative force error of 10$^{-5}$. The time step of MD simulation is 0.5 fs. The Nose Hoover thermostat was used to control the temperature of water molecules at 300K. MD simulations run 3 million timesteps and ~3000 water molecules are eventually adsorbed in MOF-801.

# References

1. Farha, O. K. *et al.* De novo synthesis of a metal-organic framework material featuring ultrahigh surface area and gas storage capacities. *Nature Chem* 2, 944–948 (2010).

2. Farha, O. K. *et al.* Metal-organic framework materials with ultrahigh surface areas: Is the sky the limit? *J. Am. Chem. Soc.* 134, 15016–15021 (2012).

3. Furukawa, H. *et al.* Ultrahigh porosity in metal-organic frameworks. *Science* 329, 424–428 (2010).

4. An, J. *et al.* Metal-adeninate vertices for the construction of an exceptionally porous metal-organic framework. *Nat Commun* 3, 604 (2012).

5. Deng, H. *et al.* Large-pore apertures in a series of metal-organic frameworks. *Science* 336, 1018–1023 (2012).

6. Sun, C.-Y., Qin, C., Wang, X.-L. & Su, Z.-M. Metal-organic frameworks as potential drug delivery systems. *Expert Opin Drug Deliv* 10, 89–101 (2013).

7. Tian, T. *et al.* A sol-gel monolithic metal-organic framework with enhanced methane uptake. *Nat. Mater.* 17, 174–179 (2018).

8. Connolly, B. M. *et al.* Tuning porosity in macroscopic monolithic metal-organic frameworks for exceptional natural gas storage. *Nat Commun* 10, 2345 (2019).

9. Chen, Z. *et al.* Fine-Tuning a Robust Metal-organic framework toward enhanced clean energy gas storage. *J. Am. Chem. Soc.* 143, 18838–18843 (2021).

10. Fan, H. *et al.* MOF-in-COF molecular sieving membrane for selective hydrogen separation. *Nat Commun* 12, 38 (2021).

11. Niu, Z. *et al.* A MOF-based ultra-strong acetylene nano-trap for highly efficient $C_2H_2/CO_2$ separation. *Angewandte Chemie* 133, 5343–5348 (2021).

12. Gu, X.-W. *et al.* Immobilization of lewis basic sites into a stable ethane-selective MOF enabling one-step separation of ethylene from a ternary mixture. *J. Am. Chem. Soc.* 144, 2614–2623 (2022).

13. Zhang, Y. *et al.* Constructing free standing metal organic framework MIL-53 membrane based on anodized aluminum oxide precursor. *Sci Rep* 4, 4947 (2014).

14. Ryu, U. *et al.* Recent advances in process engineering and upcoming applications of metal-organic frameworks. *Coordination Chemistry Reviews* 426, 213544 (2021).

15. Ren, J., Langmi, H. W., North, B. C. & Mathe, M. Review on processing of metal-organic framework (MOF) materials towards system integration for hydrogen storage. *International Journal of Energy Research* 39, 607–620 (2015).

16. Tian, T., Velazquez-Garcia, J., Bennett, T. D. & Fairen-Jimenez, D. Mechanically and chemically robust ZIF-8 monoliths with high volumetric adsorption capacity. *J. Mater. Chem. A* 3, 2999–3005 (2015).

17. Peng, Y. *et al.* Methane storage in metal-organic frameworks: Current records, surprise findings, and challenges. *J. Am. Chem. Soc.* 135, 11887–11894 (2013).

18. Bazer-Bachi, D., Assié, L., Lecocq, V., Harbuzaru, B. & Falk, V. Towards industrial use of metal-organic framework: Impact of shaping on the MOF properties. *Powder Technology* 255, 52–59 (2014).

19. Zheng, J. *et al.* Shaping of ultrahigh-loading MOF pellet with a strongly anti-tearing binder

for gas separation and storage. *Chemical Engineering Journal* 354, 1075–1082 (2018).

20. Cousin-Saint-Remi, J. *et al.* Highly robust MOF polymeric beads with a controllable size for molecular separations. *ACS Appl. Mater. Interfaces* 11, 13694–13703 (2019).

21. Çamur, C. *et al.* Monolithic zirconium-based metal-organic frameworks for energy-efficient water adsorption applications. *Advanced Materials* 35, 2209104 (2023).

22. Vilela, S. M. F. *et al.* A robust monolithic metal-organic framework with hierarchical porosity. *Chem. Commun.* 54, 13088–13091 (2018).

23. Casco, M. E. *et al.* High-pressure methane storage in porous materials: Are carbon materials in the pole position? *Chem. Mater.* 27, 959–964 (2015).

24. Madden, D. G. *et al.* Densified HKUST-1 monoliths as a route to high volumetric and gravimetric hydrogen storage capacity. *J. Am. Chem. Soc.* 144, 13729–13739 (2022).

25. Kim, H. *et al.* Water harvesting from air with metal-organic frameworks powered by natural sunlight. *Science* 356, 430–434 (2017).

26. Kim, H. *et al.* Adsorption-based atmospheric water harvesting device for arid climates. *Nat Commun* 9, 1191 (2018).

27. Wang, G., Li, Y., Qiu, H., Yan, H. & Zhou, Y. High-performance and wide relative humidity passive evaporative cooling utilizing atmospheric water. *Droplet* 2, e32 (2023).

28. Wang, G., Fan, H., Li, J., Li, Z. & Zhou, Y. Direct observation of tunable thermal conductance at solid/porous crystalline solid interfaces induced by water adsorbates. *Nat Commun* 15, 2304 (2024).

29. Bueken, B. *et al.* Gel-based morphological design of zirconium metal-organic frameworks. *Chem. Sci.* 8, 3939–3948 (2017).

30. Furukawa, H. *et al.* Water adsorption in porous metal-organic frameworks and related materials. *J. Am. Chem. Soc.* 136, 4369–4381 (2014).

31. Li, Q. *et al.* High efficient water/ethanol separation by a mixed matrix membrane incorporating MOF filler with high water adsorption capacity. *Journal of Membrane Science* 544, 68–78 (2017).

32. Fuchs, A. *et al.* Single crystals heterogeneity impacts the intrinsic and extrinsic properties of metal-organic frameworks. *Advanced Materials* 34, 2104530 (2022).

33. Butova, V. V., Pankin, I. A., Burachevskaya, O. A., Vetlitsyna-Novikova, K. S. & Soldatov, A. V. New fast synthesis of MOF-801 for water and hydrogen storage: Modulator effect and recycling options. *Inorganica Chimica Acta* 514, 120025 (2021).

34. Sun, J. *et al.* MOF-801 incorporated PEBA mixed-matrix composite membranes for $CO_2$ capture. *Separation and Purification Technology* 217, 229–239 (2019).

35. Park, S., Baker, J. O., Himmel, M. E., Parilla, P. A. & Johnson, D. K. Cellulose crystallinity index: measurement techniques and their impact on interpreting cellulase performance. *Biotechnol Biofuels* 3, 10 (2010).

36. Cai, G., Yan, P., Zhang, L., Zhou, H.-C. & Jiang, H.-L. Metal-organic framework-based hierarchically porous materials: synthesis and applications. *Chem. Rev.* 121, 12278–12326 (2021).

37. Chen, Z. *et al.* Study of the scale-up effect on the water sorption performance of MOF materials. *ACS Mater. Au* 3, 43–54 (2023).

38. Wang, C. *et al.* Remarkable adsorption performance of MOF-199 derived porous carbons for benzene vapor. *Environmental Research* 184, 109323 (2020).

39. Wang, M. & Yu, F. High-throughput screening of metal-organic frameworks for water harvesting from air. *Colloids and Surfaces A: Physicochemical and Engineering Aspects* 624, 126746 (2021).

40. Wang, X. *et al.* Benzene/toluene/water vapor adsorption and selectivity of novel C-PDA adsorbents with high uptakes of benzene and toluene. *Chemical Engineering Journal* 335, 970–978 (2018).

41. Jeremias, F., Khutia, A., Henninger, S. K. & Janiak, C. MIL-100(Al, Fe) as water adsorbents for heat transformation purposes-a promising application. *J. Mater. Chem.* 22, 10148–10151 (2012).

42. Canivet, J., Fateeva, A., Guo, Y., Coasne, B. & Farrusseng, D. Water adsorption in MOFs: fundamentals and applications. *Chem. Soc. Rev.* 43, 5594–5617 (2014).

43. Solovyeva, M. V., Gordeeva, L. G., Krieger, T. A. & Aristov, Yu. I. MOF-801 as a promising material for adsorption cooling: Equilibrium and dynamics of water adsorption. *Energy Conversion and Management* 174, 356–363 (2018).

44. Kalmutzki, M. J., Diercks, C. S. & Yaghi, O. M. Metal-organic frameworks for water harvesting from air. *Advanced Materials* 30, 1704304 (2018).

45. AbdulHalim, R. G. *et al.* A Fine-Tuned Metal-organic framework for autonomous indoor moisture control. *J. Am. Chem. Soc.* 139, 10715–10722 (2017).

46. Qin, M., Hou, P., Wu, Z. & Wang, J. Precise humidity control materials for autonomous regulation of indoor moisture. *Building and Environment* 169, 106581 (2020).

47. Fathieh, F. *et al.* Practical water production from desert air. *Sci. Adv.* 4, eaat3198 (2018).

48. Zhang, B., Zhu, Z., Wang, X., Liu, X. & Kapteijn, F. Water adsorption in MOFs: structures and applications. *Advanced Functional Materials* n/a, 2304788.

49. Datta, A. K. Porous media approaches to studying simultaneous heat and mass transfer in food processes. I: Problem formulations. *Journal of Food Engineering* 80, 80–95 (2007).

50. Datta, A. K. Porous media approaches to studying simultaneous heat and mass transfer in food processes. II: Property data and representative results. *Journal of Food Engineering* 80, 96–110 (2007).

51. Plimpton, S. Fast parallel algorithms for short-range molecular dynamics. *Journal of Computational Physics* 117, 1–19 (1995).

52. Wu, Y., Tepper, H. L. & Voth, G. A. Flexible simple point-charge water model with improved liquid-state properties. *The Journal of Chemical Physics* 124, 024503 (2006).

53. Rappe, A. K., Casewit, C. J., Colwell, K. S., Goddard, W. A. I. & Skiff, W. M. UFF, a full periodic table force field for molecular mechanics and molecular dynamics simulations. *J. Am. Chem. Soc.* 114, 10024–10035 (1992).

**Acknowledgments**

Y.Z. thanks the Equipment Competition fund (REC20EGR14), the open fund from the State Key Laboratory of Clean Energy Utilization (ZJUCEU2022009), the Frontier Technology Research for Joint Institutes with Industry Scheme FTRIS-002 and the Hong Kong SciTech Pioneers Award from the Y-LOT Foundation. Y.Z. and Z.L. thank the Research Grants Council of the Hong Kong Special Administrative Region under Grant C6020-22G. Y.Z. also thanks the Research Grants Council of the Hong Kong Special Administrative Region under Grant 260206023 and C7002-22Y, the fund from the Guangdong Natural Science Foundation under Grant No. 2024A1515011407. Y.Z. and C.T. thank the Hong Kong Environment and Conservation Fund under Grant No. 08/2022. The authors are grateful to the Materials Characterization and Preparation Facility (MCPF) of HKUST for their assistance in experimental characterizations.

**Author contributions**

Y.Z. conceived the idea and supervised the project; J.L. designed the experiments and conducted the material synthesis, characterization, and performance investigation; H.F. did the molecular dynamics simulations; J.L., G.W., F.H., and Y.Z. prepared the manuscript; all the authors reviewed and revised the manuscript.

**Supplementary information**

The online version contains available supplementary information.